\documentclass[epjCONF,columns]{svjour}
\usepackage{graphics}
\usepackage[varg]{txfonts} 
\usepackage[latin1]{inputenc}

\usepackage{graphicx}

\session-title{Hadron Collider Physics symposium (HCP-2011)}
\usepackage{lineno}

\begin{document}
%
\title{Search for charged Higgs bosons in ATLAS}
\author{Daniel Pelikan, on behalf of the ATLAS Collaboration\thanks{\email{daniel.pelikan@cern.ch}}}
\institute{Uppsala University, Sweden}
\abstract{
Charged Higgs bosons are predicted in several extensions of the
Standard Model, where the Higgs sector contains more than one doublet of complex scalars, for instance in the Minimal Supersymmetric Standard Model (MSSM). For $m_{H^+} < m_{\mathrm{top}}$, the dominant production mode for
charged Higgs bosons at the LHC is via the decay $t\rightarrow bH^+$ of one of
the top quarks in $t\bar{t}$ events. We present results on the search for
such light charged Higgs bosons in the ATLAS experiment, with emphasis
on the decay $H^+ \rightarrow \tau \nu$. 
} 
\maketitle

\section{Introduction}
\label{intro}
\begin{figure}[htbp]
\centering
 \includegraphics[width=0.55\linewidth]{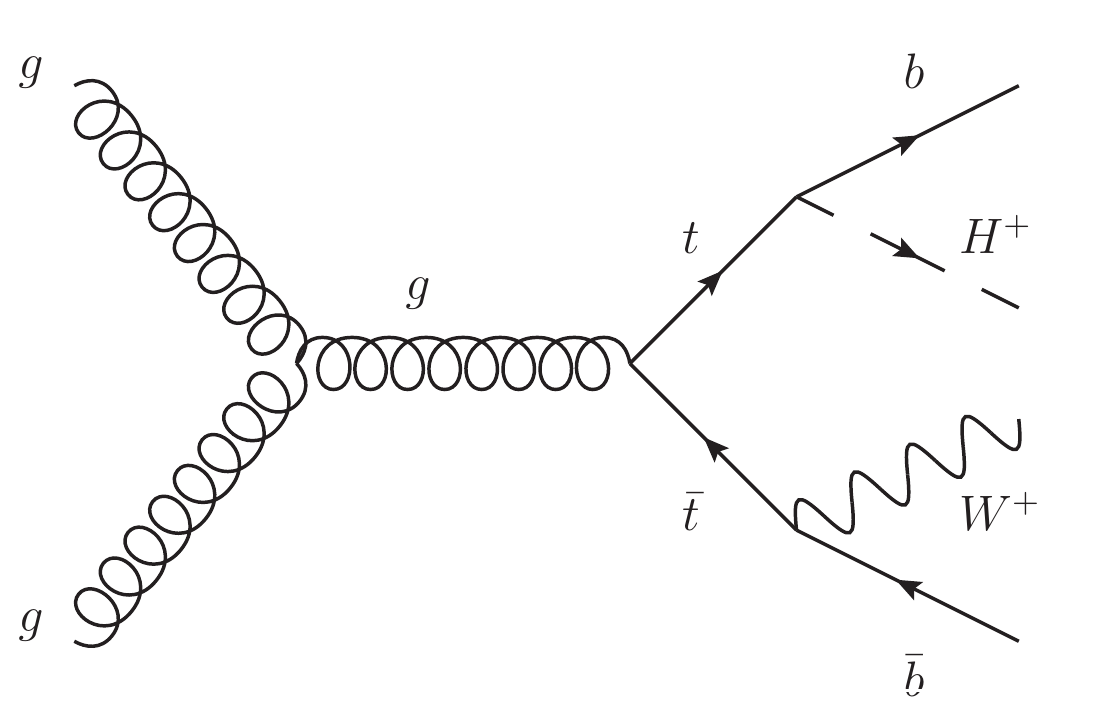}
\caption{Feynman diagram of the charged Higgs boson production in $t\bar{t}$ events.}
\label{FineMan}
\end{figure}

One major task for the Large Hadron Collider (LHC) and the ATLAS experiment is the
exploration of the Higgs sector in the Standard Model (SM) and beyond it. 
Charged Higgs bosons, $H^+$ and $H^-$, are predicted by several extensions of the SM, such
as models containing Higgs triplets and Two-Higgs-Doublet Models (2HDM) \cite{ChHiggs}. Electroweak symmetry breaking through two complex Higgs doublets lead to five
physical states, out of which two are charged. The observation of charged Higgs bosons would clearly indicate physics beyond the SM. Here the type-II 2HDM is considered, 
which is also the Higgs sector of the Minimal Supersymmetric Standard Model (MSSM) \cite{MSSM}. 
For values of $\tan\beta$ (the ratio of the vacuum expectation values 
of the two Higgs doublets) larger than 3, typically, charged Higgs bosons decay mainly via $H^+\rightarrow\tau\nu$ \cite{ChHiggsDec}. 
In the following, a branching ratio $\mathcal{B}(H^+\rightarrow\tau\nu)=1$ is considered.\\
\\
The searches for the charged Higgs boson in $t\bar{t}$ events (see Fig. \ref{FineMan}) with one or two light charged leptons (i.e. electrons or muons) in the final state \cite{OwnNote} and in the case where both the $\tau$ and the $W$ boson decay hadronically ($\tau+$jets channel) \cite{TauHatNote}
are described. $1.03$ fb$^{-1}$ of data from proton-proton collisions, at a centre of mass energy of $\sqrt{s}=7$ TeV, collected with the ATLAS experiment in 2011, are analysed. 
Upper limits for $\mathcal{B}(t\rightarrow bH^\pm)$ and the charged Higgs boson production in the $m_{H^+}-\tan{\beta}$ plane ($m_H^{\mathrm{max}}$ scenario \cite{mHmax}) are derived.
\section{The $\tau$+jets Final State}
This study describes the search for the charged Higgs boson in $t\bar{t}$ events which decay into $[H^+b] [W^-\bar{b}]$ and then into $[(\tau^+_{\mathrm{had}}+\nu)b][q\bar{q}' \bar{b}]$. 
For such $\tau$+jets events, the transverse mass $m_\mathrm{T}^\tau$ is defined as:
\begin{equation}\label{mtTau}
 m_\mathrm{T}^\tau=\sqrt{2p_\mathrm{T}^{\tau}E_\mathrm{T}^{\mathrm{miss}}(1-\cos\Delta\phi)},
\end{equation}
where $\Delta\phi$ is the azimuthal angle between the $\tau$-jet and the missing energy direction.
The main sources of background are $t\bar{t}$ and single top-quark production, $W$+jets and multi-jet events. All these backgrounds are determined with data-driven methods.
\subsection{Electron-to-$\tau$ misidentification probability with a tag-and-probe method}
Background events where an electron is misidentified as a hadronically decaying $\tau$ are estimated from a control sample using a data-driven tag-and-probe method. In order to determine the misidentification probability, $Z/\gamma^*$ events in collision data are used. The result in data is compared to Monte Carlo (MC) and the ratio is used as a scale factor in MC. 
\subsection{Jet-to-$\tau$ misidentification probability from photon+jets events}
In order to determine the probability for a jet to be misidentified as a hadronically decaying $\tau$, a $\gamma$+jet control sample is used. Like jets from the hard processes in the dominant $t\bar{t}$ background, jets in the control sample originate mainly from quarks. A measurement of the probability for a jet to be misidentified as a hadronically decaying $\tau$ is performed in data and is used to predict the yield of jet-to-$\tau$ misidentification events from the most important SM backgrounds. 
\subsection{Multi-jet background estimate}
The expectations in MC for multi-jet events have large uncertainties, for that reason this background is estimated from data.
This is achieved by fitting its $E_\mathrm{T}^{\mathrm{miss}}$ shape to data. For the shape study a control region is defined where the $\tau$ identification and the $b$-tagging are inverted.

 \subsection{Embedding method}
 The embedding method is used to estimate the background from true $\tau$ jets. Control samples of single-top, $W$+jets, and $t\bar{t}$ events with a muon in data are collected and the detector signature of this muon is replaced with that of a simulated hadronically decaying $\tau$. Then, the reconstruction is re-applied on the new hybrid event. 
\subsection{Results} 
\begin{figure}
\centering
 \includegraphics[width=0.75\linewidth]{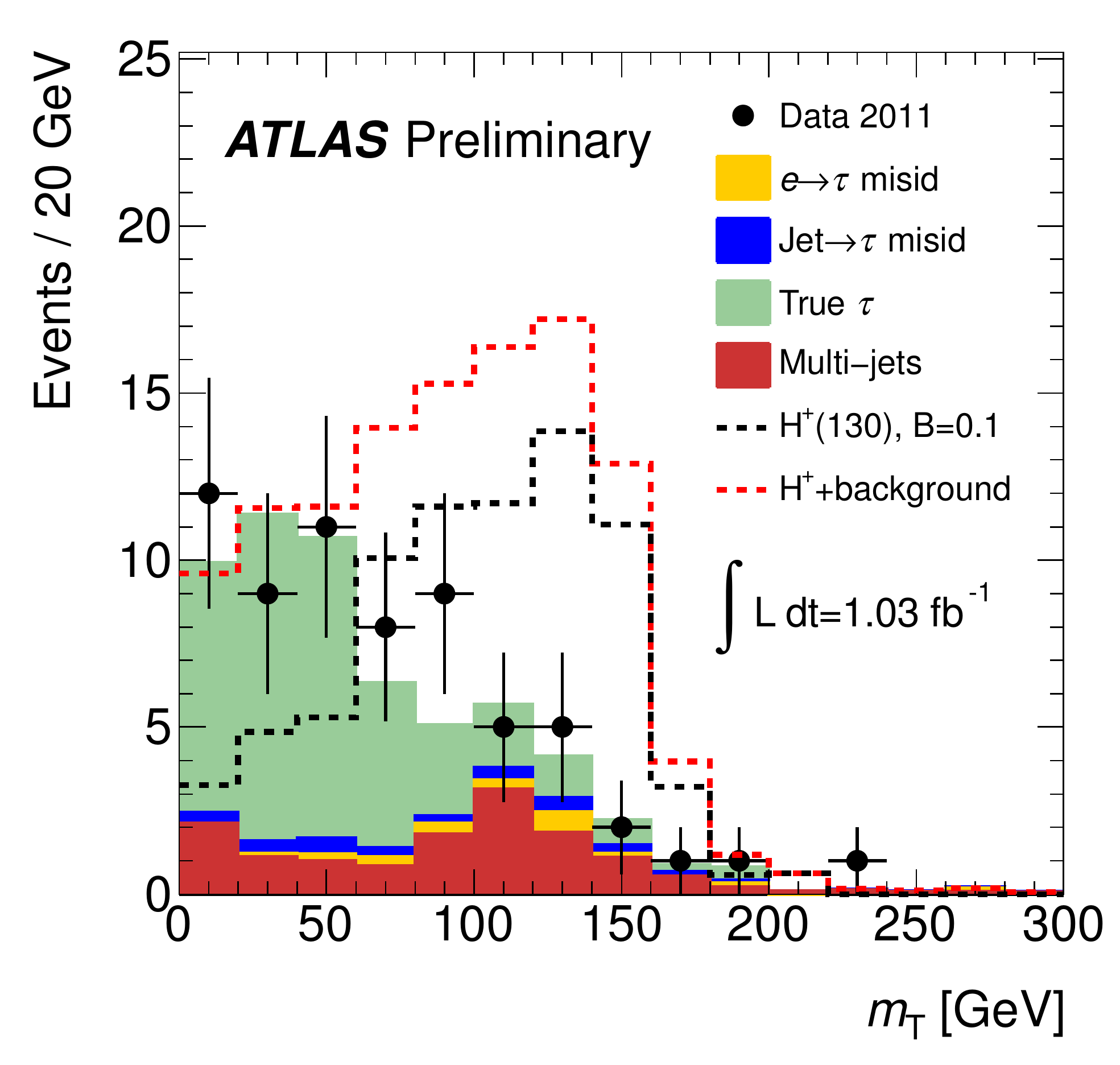}
\caption{$m_\mathrm{T}$ distribution after event selection.}
\label{FigMtHad}
\end{figure}

\begin{figure}
\centering
 \includegraphics[width=0.75\linewidth]{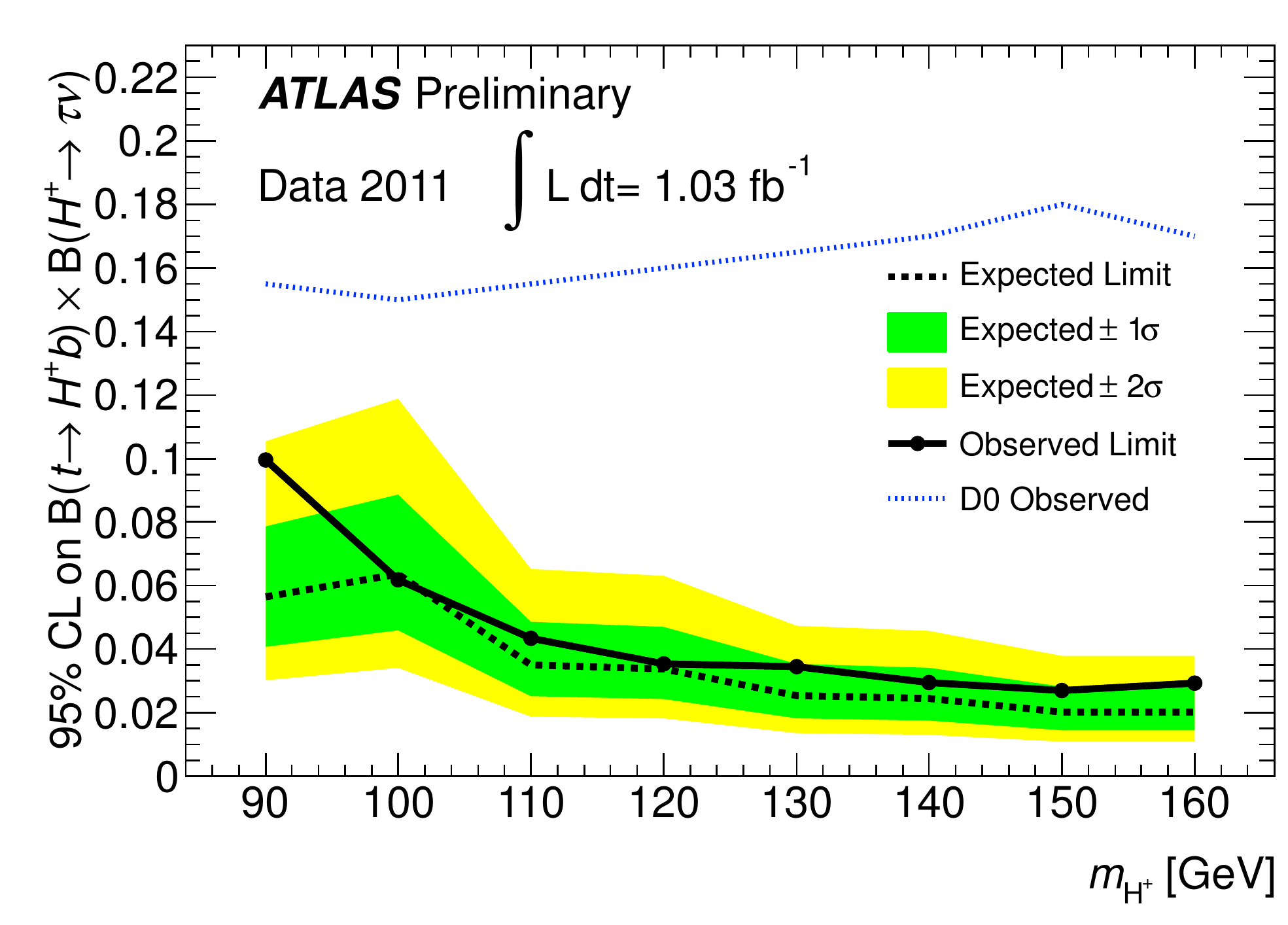}
\caption{Expected and observed limit with 95\% C.L. of $\mathcal{B}(t\rightarrow~H^+b)$ from $\tau$+jets events as a function of $m_H^+$ and the comparison with the D0 experiment.}
\label{HadLimB}
\end{figure}

\begin{figure}
\centering
 \includegraphics[width=0.75\linewidth]{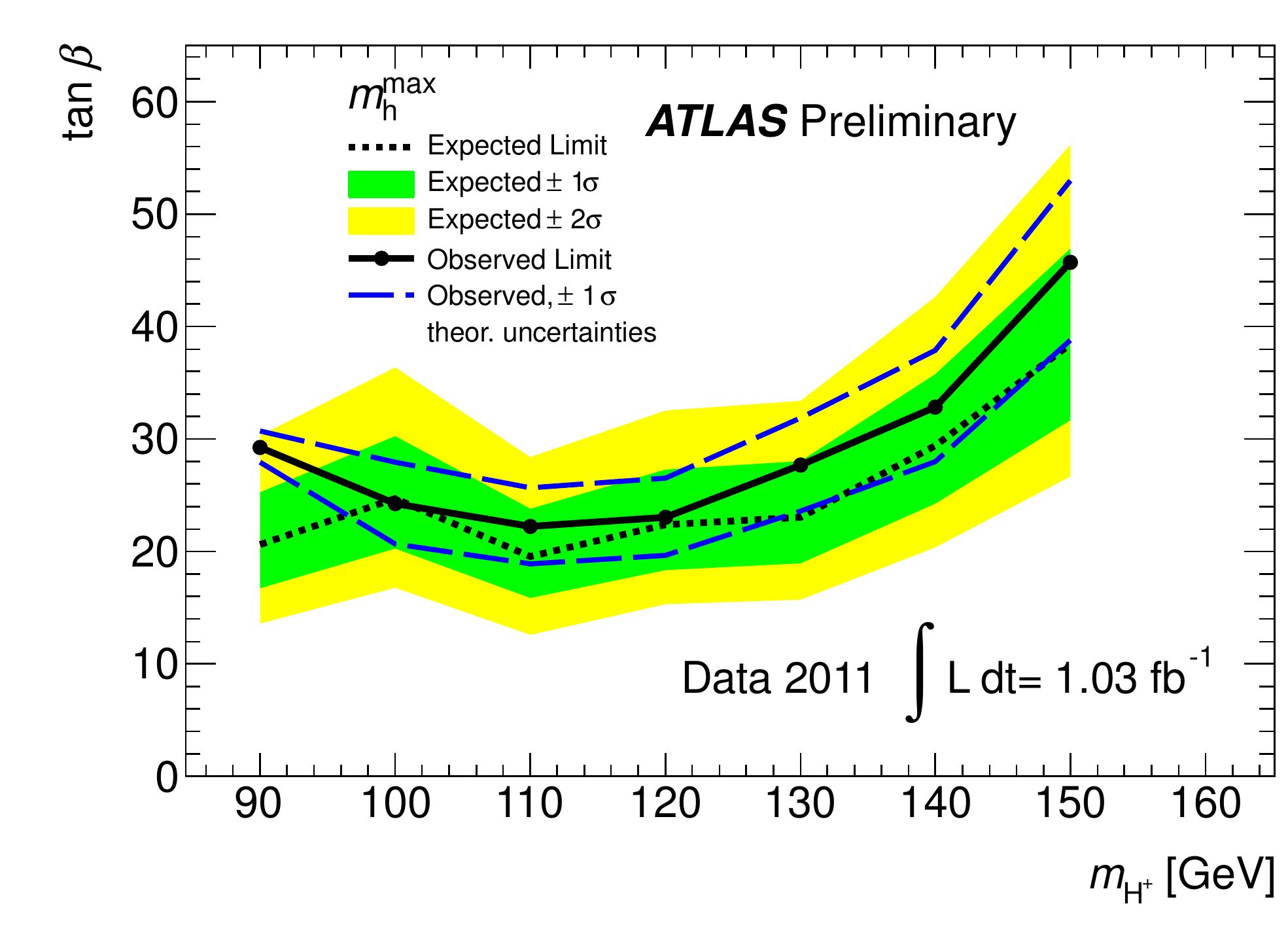}
\caption{Limit from $\tau$+jets events for $H^+$ production in the $m_{H^+}-\tan\beta$ plane ($m_H^{\mathrm{max}}$ scenario).}
\label{HadLimProd}
\end{figure}

In Fig. \ref{FigMtHad} the $m_\mathrm{T}$ distribution with all backgrounds after all cuts is shown.
%
%
Assuming $\mathcal{B}(H^+\rightarrow\tau\nu)=1$, upper limits are extracted on the branching ratio $B\equiv\mathcal{B}(t\rightarrow b H^+)$ as a function of the charged Higgs boson mass.
A profile likelihood statistical analysis is performed in order to set upper limits with a 95\% Confidence Level (C.L.) on the branching ratio (Fig.~\ref{HadLimB}) and the $H^+$ production in the $m_H^{\mathrm{max}}$ scenario (Fig.~\ref{HadLimProd}). A limit on the branching ratio below 10\% is reached.

\section{Single- and Di-Lepton Final States}
In this channel, the $\tau$ from the charged Higgs boson decays leptonically, the $W$ boson on the other side decays either hadronically (single-lepton) or leptonically (di-lepton).
To search for charged Higgs bosons in the single- 
and di-lepton channels, discriminating variables are used, that allow a distinction between leptons from $\tau\rightarrow l \nu_l\nu_{\tau}$ (e.g. in decays of $W$ or charged Higgs bosons) or leptons coming directly from the $W$ boson decay.
The invariant mass $m_{bl}$ of a $b$-quark and a light charged lepton $l$ (electron or muon) can be used, or more conveniently, the variable $\cos\theta^*_l$ 
defined as:
\begin{equation}
  \cos\theta_{l}^* = \frac{2m_{bl}^2}{m_{\mathrm{top}}^2-m_W^2}-1 \simeq 
\frac{4\,p^b \cdot p^l}{m_{\mathrm{top}}^2-m_W^2}-1,\label{CosEq} 
\end{equation}
where $p^b$ and $p^l$ are the four-momenta of the $b$-quark and the lepton, $m_{\mathrm{top}}$ and $m_W$ are the mass of the top-quark and the $W$ boson.
\\
Other discriminating variables are two new transverse mass observables \cite{Offer}. 
In single-lepton events, the \textit{charged Higgs boson transverse mass} 
$m_\mathrm{T}^H$ is computed as follows:
\begin{equation}
 (m_\mathrm{T}^H)^2 = \left( \sqrt{m_\mathrm{top}^2+(\vec p_\mathrm{T}^l+ \vec p_\mathrm{T}^b+\vec p_\mathrm{T}^\mathrm{miss})^2}- p_\mathrm{T}^b\right)^2-\left(\vec p_\mathrm{T}^l+\vec p_\mathrm{T}^\mathrm{miss}\right)^2
\end{equation}
In the case of di-lepton events, the \textit{generalised charged Higgs boson transverse mass} $m_\mathrm{T2}^H$ is obtained as follows:
\begin{equation}
 m_\mathrm{T2}^H = 
\mathop {\max }\limits_{\left\{ \scriptstyle constraints\right\} } {\kern 1pt} [m_\mathrm{T}^H(\vec p_\mathrm{T}^{\hspace{1mm}H^+})],
\end{equation}
where some kinematical constraints (see eq. (9) in \cite{Offer}) must be fulfilled and with the definition:
\begin{equation}
 m_\mathrm{T}^H(\vec p_\mathrm{T}^{\hspace{1mm}H^+}) )^2  = 
\left(\sqrt {m_{\mathrm{top}}^2  + (\vec p_\mathrm{T}^{\hspace*{1mm}H^+} + 
\vec p_\mathrm{T}^{\hspace*{1mm}b})^2 } - 
p_\mathrm{T}^b \right)^2  - \left(\vec p_\mathrm{T}^{\hspace*{1mm}H^+}\right)^2.
\end{equation}
The multi-jet background is estimated using data-driven methods, the other backgrounds are taken from simulation.

\subsection{Single-Lepton Events}
\begin{figure}[]
\centering
\includegraphics[width=0.75\linewidth]{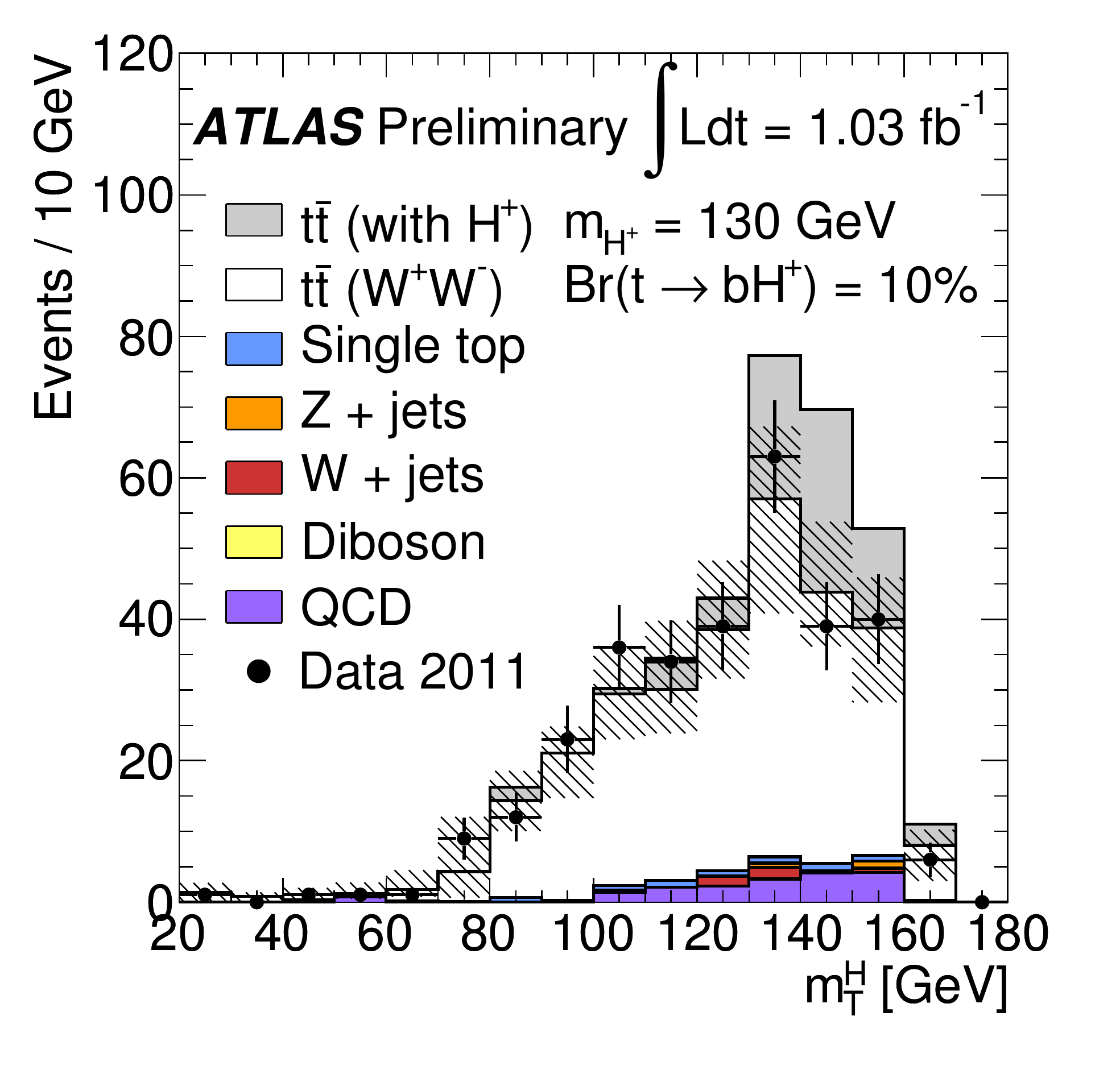}
\caption{$m_\mathrm{T}^H$ distribution with $\cos\theta^*_l<-0.6$, $m_\mathrm{T}^W<60$~GeV.}
\label{MTHiggs}
\end{figure}

In order to select single-lepton $t\bar{t}$ events, the following 
selection cuts are applied:
\begin{itemize}
 \item exactly one trigger-matched lepton with $E_\mathrm{T}>25$ GeV (electron) or $p_\mathrm{T}>20$ GeV (muon),
\item at least four jets with $p_\mathrm{T}>20$ GeV and $|\eta|<2.5$, including exactly two b-jets,
\item \begin{tabbing}
\kill
  $E_\mathrm{T}^{\mathrm{miss}}>40$ GeV $\quad\quad\quad\quad\quad$ \=if $|\phi_{l,\mathrm{miss}}|\geq\pi/6$,\\
      $E_\mathrm{T}^{\mathrm{miss}}\times | \sin(\phi_{l,\mathrm{miss}})|>20$ \> if $|\phi_{l,\mathrm{miss}}|<\pi/6$.     
      \end{tabbing}
\end{itemize}
After having assigned two light-quark jets and one $b$-jet 
to the hadronically decaying top-quark candidate, the other $b$-jet 
is associated to the lepton and the charged Higgs boson transverse 
mass is reconstructed, see Fig. \ref{MTHiggs}. To enhance the sample with leptonically decaying $\tau$s, additional cuts on $\cos\theta^*_l<-0.6$ and $m_\mathrm{T}^W<60$ GeV are applied. $m_\mathrm{T}^W$ is defined as in Eq.~(\ref{mtTau}), where $\tau$ is replaced by the lepton.

\subsection{Di-Lepton Events}
To select di-lepton $t\bar{t}$ events the following cuts are applied:
\begin{itemize}
 \item exactly two oppositely charged leptons, including at least one matched to the trigger
with $E_\mathrm{T} > 25$ GeV (electron) or $p_\mathrm{T} > 20$ GeV (muon),
\item at least two jets with $p_\mathrm{T} > 20$ GeV and $|\eta| < 2.5$, including exactly two $b$-tagged jets,
\item for $ee$ and $\mu\mu$ events, the invariant mass $m_{ll}$ must exceed $15$ GeV and must satisfy
$|m_{ll} - m_Z| > 10$ GeV, together with $E^{\mathrm{miss}}_\mathrm{T}> 40$ GeV,
\item for $e\mu$ events, the scalar sum of the transverse energies of the selected leptons and jets is $\sum E_\mathrm{T} > 130$ GeV.
\end{itemize}

\begin{figure}[]
\centering
\includegraphics[width=0.75\linewidth]{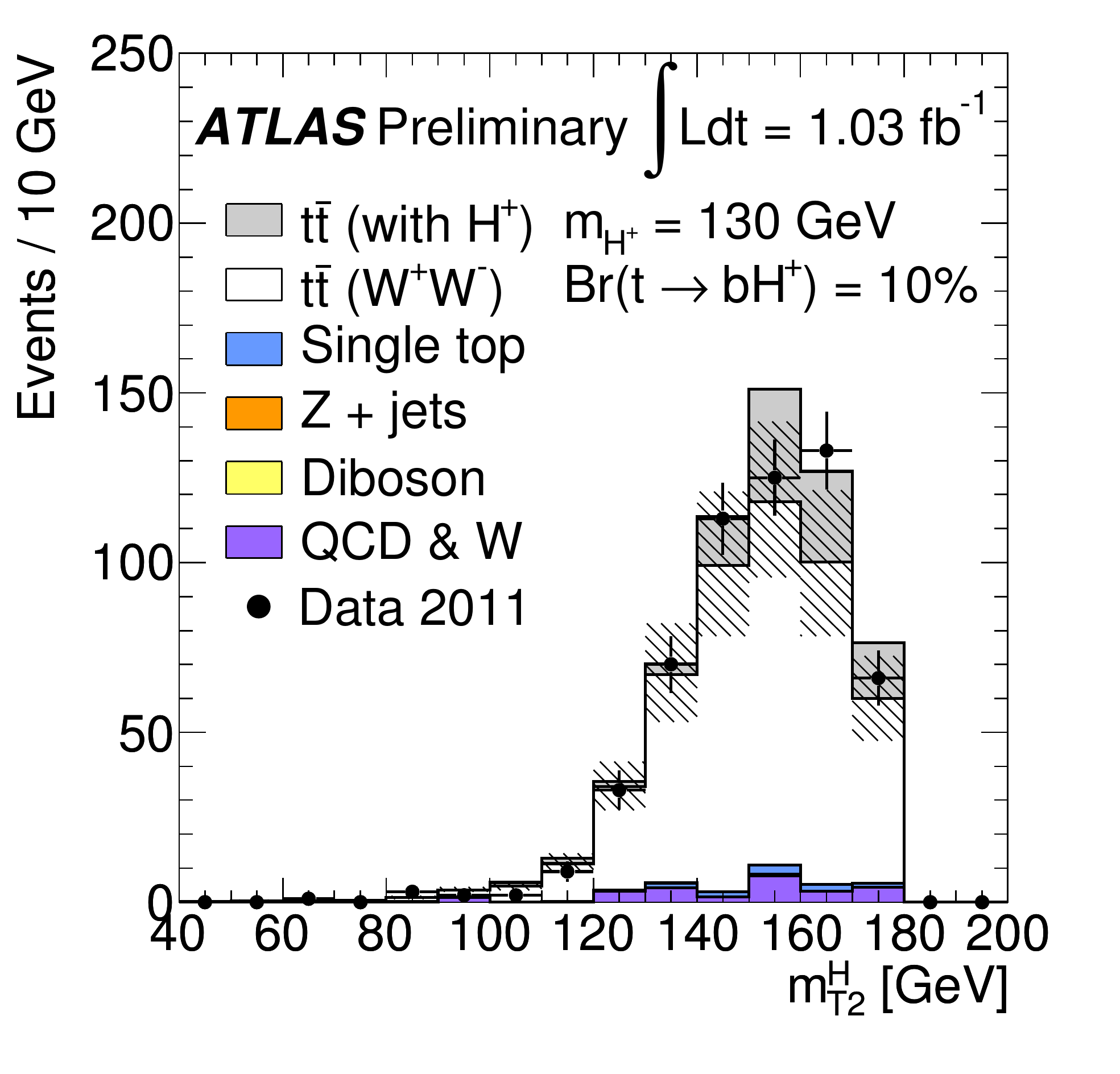}
\caption{$m_\mathrm{T2}^H$ distribution for events with $\cos\theta^*_l<-0.6$.}
\label{MT2Higgs}
\end{figure}

After assignment of the two leptons 
and the two $b$-jets to their correct parents, based on $\cos\theta^*_l$ and the 
distance between each lepton and $b$-jet, the generalised charged Higgs 
boson mass is reconstructed, see Fig. \ref{MT2Higgs}.

%
%
%
%
%

\subsection{Limits on the Branching Ratio of $t\rightarrow b H^+$}
\begin{figure}
 \centering
\includegraphics[width=0.75\linewidth]{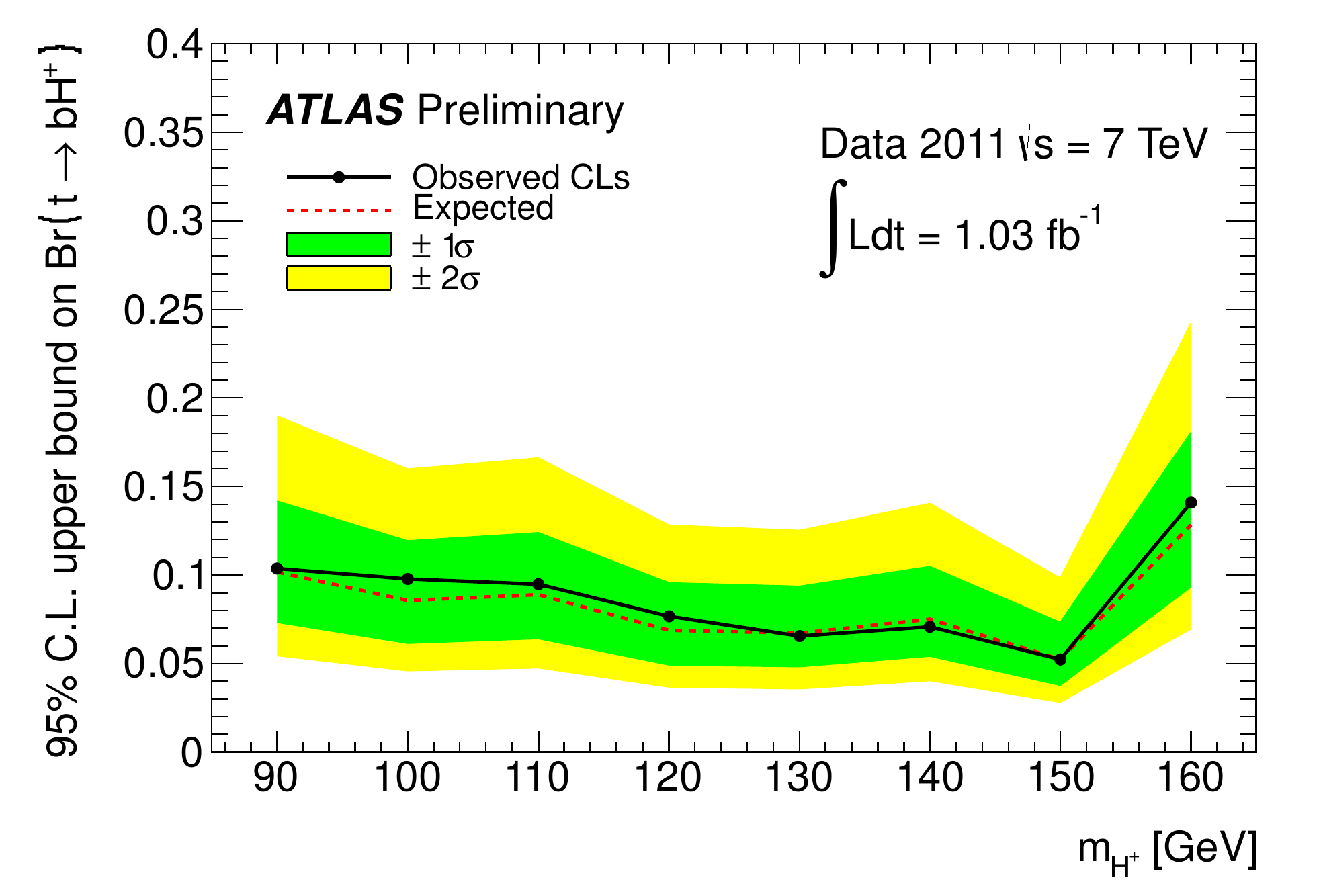}
\caption{ Expected and observed combined upper limits of $\mathcal{B}(t\rightarrow~H^+b)$ with 95\% C.L. as a function of $m_{H^+}$.}
\label{BrLimSiDi}
\end{figure}

\begin{figure}
 \centering
\includegraphics[width=0.75\linewidth]{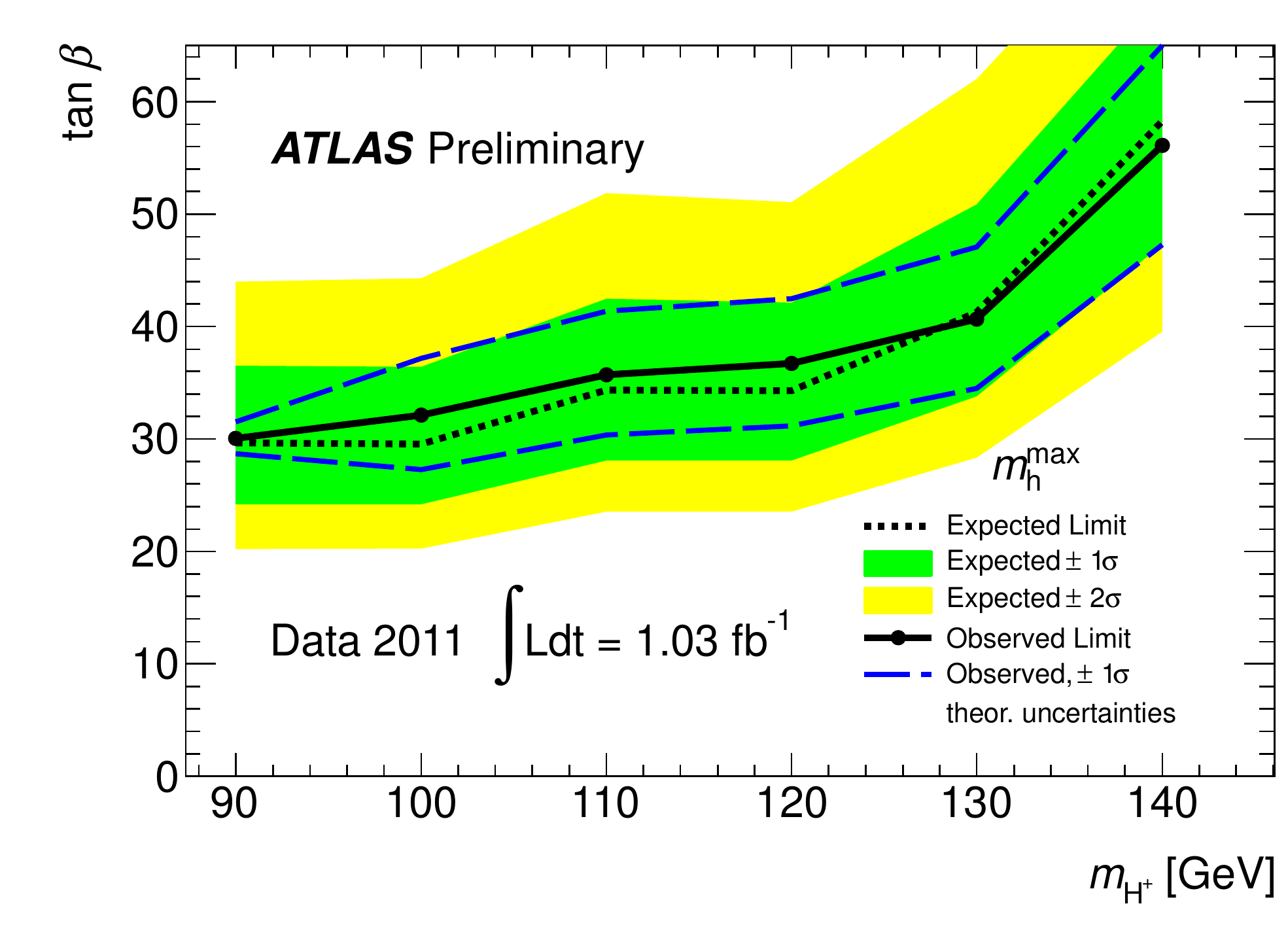}
\caption{ Combined limits for $H^+$ production in the $m_{H^+}$-$\tan\beta$ plane ($m_H^\mathrm{max}$ scenario).}
\label{ProdLimSiDi}
\end{figure}

The combined limits, from single- and di-lepton events, with a C.L. of 95\% for the branching ratio $B\equiv\mathcal{B}(t\rightarrow~b~H^+)$ and the limit on the production for the charged Higgs boson are shown in Fig. \ref{BrLimSiDi} and \ref{ProdLimSiDi}.


\section{Conclusion}
Results for the search of charged Higgs bosons in ATLAS are presented using 1.03 fb$^{-1}$ of $pp$ collision data at 7~TeV. Upper limits for the branching ratio $\mathcal{B}(t\rightarrow b H^+)$ between 3\% and 10\% for charged Higgs boson masses in the range of 90 GeV $< m_H^+ <$ 160 GeV have been derived, with the analysis of more ATLAS data improvements are expected.

%
%

\end{document}